\documentclass[
  aps,prl,reprint,amsmath,amssymb,
  superscriptaddress,floatfix,
  longbibliography,nofootinbib,
  nobalancelastpage
]{revtex4-2}

\usepackage{graphicx}
\usepackage{orcidlink}
\usepackage{bm}
\usepackage{booktabs}
\usepackage{xcolor}
\usepackage{hyperref}
\hypersetup{
    colorlinks=true,
    citecolor=blue,
    linkcolor=blue,
    urlcolor=blue
}
\graphicspath{{figures/}{PDFs_of_Scalar_meson_in_Schwinger_model/figures/}}

\newcommand{\ii}{\mathrm{i}}
\newcommand{\dd}{\mathrm{d}}
\newcommand{\vac}{\Omega}
\newcommand{\Wline}{\mathcal W}
\newcommand{\ket}[1]{\lvert #1\rangle}
\newcommand{\bra}[1]{\langle #1\rvert}
\definecolor{unrevised}{RGB}{200,112,255}

\newcommand{\prlsection}[1]{%
  \par\noindent\textit{#1---}\enspace\ignorespaces}

\makeatletter
\DeclareRobustCommand{\equalcontrib}{%
  \texorpdfstring{\frontmatter@footnote{These authors contributed equally to this work.}}{}%
}
\makeatother

\begin{document}

\title{Probing the inner structure of bound states}

\author{Yahui Chai \orcidlink{0000-0002-3404-6096}\equalcontrib}
\email{yahui.chai@riken.jp}
\affiliation{RIKEN Center for Interdisciplinary Theoretical and Mathematical Sciences (iTHEMS),
 Wako 351-0198, Japan}
 \affiliation{RIKEN Center for Computational Science (R-CCS), Kobe 650-0047, Japan}
\author{Yibin Guo \orcidlink{0000-0003-0435-1476}\equalcontrib}
\email{yibin.guo@uni-saarland.de}
\affiliation{Department of Physics and Center for Quantum Technologies (QuTe), Saarland University, 66123 Saarbrücken, Germany}
\author{Karl Jansen \orcidlink{0000-0002-1574-7591}}
\email{karl.jansen@desy.de}
\affiliation{Computation-Based Science and Technology Research Center, The Cyprus Institute, 20 Kavafi Street, 2121 Nicosia, Cyprus}

\begin{abstract}
We show how constructing a composite particle eigenstate reveals its internal structure. Using quantum subspace expansion (QSE), we optimize states with different sets of operators and use their fidelities with the full QSE state to measure the importance of each added operator type. Taking the scalar meson of the massive Schwinger model as a benchmark, we find that fermion bilinears suffice at large fermion masses. At small masses, two classes of four-fermion operators play important but very different roles: the four fermion operators with two fermions and 2 anti-fermions at small distance give little improvement. In contrast, spatially extended meson--meson operators produce a large gain in state fidelity.  Turning to observables, we find that the scalar charge $g_S$ is much more sensitive to the meson--meson operators than e.g. the meson mass. The parton distribution function (PDF) also changes from symmetric to asymmetric about the momentum fraction $1/2$, along with the growing importance of meson--meson operators. These results establish QSE as a tool for probing the internal structure of composite particles and connecting it to observable properties.
\end{abstract}

\maketitle

\prlsection{Introduction}
Understanding how constituent particles form bound states and what are their properties is a general goal in different areas of physics. For example, in QCD so-called parton distribution functions can be measured in large scale experiments and also theoretically analyzed. This provides important information on quarks and gluons, and how they form hadrons. From such analyses we have learned that the proton contains three valence quarks, as well as sea quarks and gluons. Also, their spin and momentum distributions remain under study~\cite{Alexandrou2017,Burkardt2000} employing non-perturbative lattice simulations. However, there are open problems such as scalar mesons which have been described as quark--antiquark states, multiquark states, or meson--meson states~\cite{Jaffe1977,Pelaez2016,Dudek2016,Guo2018}. Deciding which of these pictures is correct, requires to have first principle computations of parton distributions and fragmentation functions to probe exotic-hadron structure~\cite{Hirai2008,Han2024} which remains  challenging for lattice calculations. 

While lattice QCD studies were very successful in calculating, e.g. hadron spectra using Euclidean correlation functions~\cite{Dudek2016,Alexandrou2017}, extracting structure-sensitive observables, such as scalar charges, is difficult because of excited-state contamination and noisy disconnected diagrams~\cite{Gupta2021}. In addition, the correlators defining PDF are not directly accessible in Euclidean simulations~\cite{Ji2013}, because they are defined on the light cone. It would therefore be very natural to use a Hamiltonian formulation with a Minkowski space-time signature. Indeed, Hamiltonian methods instead correspond to correlation functions to be evaluated through real-time evolution~\cite{Li2022,LiXing2026}. Tensor-network calculations have used this approach to obtain PDFs in the Nambu--Jona-Lasinio and Schwinger models~\cite{Kang2025,Banuls2026,Grieninger2024,Grieninger2026}.

In our work we take this idea to access the hadron state --and hence its constituent composition-- directly, using quantum subspace expansion (QSE)~\cite{McClean2017,Chai2025}. While we use the Schwinger model for this purpose, we believe that the employed methods can also be used in many other models which have bound states. To be more concrete, we construct approximate eigenstates by acting with an operator basis adapted to the states we want to explore. These operators act on the interacting vacuum and we obtain the approximate eigenvalues by solving a generalized eigenvalue problem within the operator subspace. As a measure of which operators have to be added to a subspace we use the fidelities between independently optimized states in different operator subspaces. From these fidelities we can quantify the improvement from each added operator class and identify its relative importance within the employed basis.

\begin{figure}[!t]
 \centering
 \includegraphics[width=\linewidth]{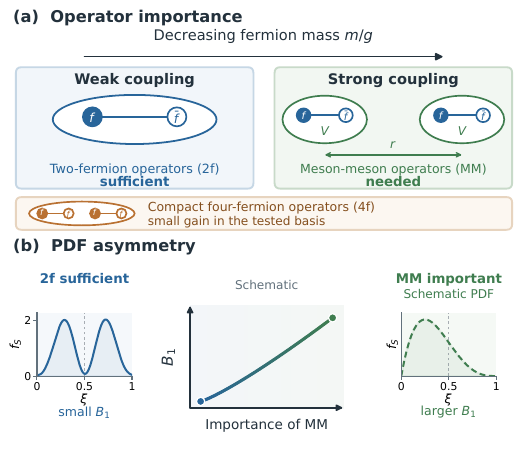}
 \caption{\textbf{Internal structure of the scalar meson and PDF asymmetry.}
 (a) Operator importance from weak to strong coupling as fermion mass $m/g$ decreases. At large $m/g$, fermion bilinears (2f) suffice to reconstruct the scalar meson. At small $m/g$, meson--meson (MM)
operators become essential, while compact four-fermion (4f) operators give little improvement within the tested basis. Each MM operator combines two vector-meson operators $V$ with center separation $r$.
 (b) Increasing PDF asymmetry accompanies the growing importance of MM operators. The asymmetry can be measures by the quantity $B_1$, which vanishes for a PDF symmetric about $\xi=1/2$. The left PDF is the real part of the full QSE result at $m_{\rm lat}/g=1$. The central $B_1$ curve and the dashed PDF on right are schematic. Gray dashed lines mark $\xi=1/2$.}
\label{fig:scalar-overview}
\end{figure}

In analogy to QCD we choose the scalar meson of the Schwinger model as the representative particle to apply this approach using tensor networks. As Figure~\ref{fig:scalar-overview}(a) summarizes, at large fermion masses, fermion bilinears reproduce the full QSE state accurately. At small masses, we find that two classes of four-fermion operators play important but very different roles: the compact four-fermion operators combines two fermions and two antifermions within a short spatial interval, give little improvement; whereas meson--meson operators consist of two spatially separated vector mesons, substantially increase the state fidelity and hence are very relevant to describe the state. This comparison highlights the importance of spatially extended two-meson operators within the basis. Interestingly, a similar conclusion with a two-meson picture was reached for light scalar mesons in QCD in earlier studies~\cite{Briceno2018, Santowsky2022}.

To examine observable consequences of the structure identified by QSE, we calculate the scalar charge $g_S$ and the PDF. Our results show that $g_S$ is more sensitive to the chosen operator basis than the meson mass, with a particularly large correction from meson--meson operators at small fermion masses. For the scalar meson, the light-front fermion--antifermion description predicts a PDF that is symmetric with respect to the momentum fraction $1/2$, in both QCD and the Schwinger model~\cite{scalar-pdf-symmetry,Harada1995}. It is therefore interesting, whether we can observe an asymmetric PDF and take any violation of symmetry as a signature of departures from the two-body description. Using real-time correlators~\cite{Banuls2026}, we find that an increasing asymmetry accompanies the growing importance of meson--meson operators as the fermion mass decreases, as sketched in Fig.~\ref{fig:scalar-overview}(b). At small fermion masses, the correlators are not converging to zero even at the largest separations available on our lattice. This makes reliable reconstruction of the full PDF difficult. We instead show how to quantify the PDF asymmetry directly from correlators at specific finite separations.

In this paper we demonstrate that the here presented approach works in practice. In particular, by combining an analysis of operator importance in state construction with calculations of the scalar charge and PDF, we connect the particle’s internal structure with its observable properties. We are confident that the methods used here can be applied to many other models where bound states are present.

\vspace{0.5cm}
\prlsection{Model}
The massive Schwinger model describes quantum electrodynamics in
$1+1$ dimensions~\cite{Schwinger1962}. In temporal gauge, $A_0=0$,
the continuum Hamiltonian is
\begin{equation}
 \mathcal{H}_{\rm cont}=\int\dd z\left[
 -\ii \bar\psi \gamma^1 (\partial_1 - \ii g A_1) \psi + m \bar\psi \psi+\frac{\mathcal E^2}{2}\right].
 \label{eq:Hcont}
\end{equation}
Here $\psi(z)$ is a two-component Dirac fermion of mass $m$, $g$ is
the gauge coupling, and $A_1$ and $\mathcal E$ are the spatial
gauge field and electric field, respectively.

We discretize the Hamiltonian using staggered
fermions~\cite{KogutSusskind1975,Susskind1977}. With open boundary conditions (OBC), the spatial gauge links can be eliminated by a gauge transformation. Gauss's law then expresses the electric field in terms of the fermion charges.
The resulting dimensionless Hamiltonian in terms of the staggered fermion operators is
\begin{align}
H={}&-\ii x\sum_{n=0}^{N-2}
 \bigl(\xi_n^\dagger\xi_{n+1}-\mathrm{h.c.}\bigr)
 +\mu\sum_{n=0}^{N-1}(-1)^n\xi_n^\dagger\xi_n\notag\\
 &+\sum_{n=0}^{N-2}\left(\sum_{j=0}^{n}Q_j\right)^2,
 \label{eq:H}
\end{align}
where $N$ is the number of staggered sites, $\xi_n^\dagger,\xi_n$ are fermion creation and annihilation operators respectively, and $Q_n=\xi_n^\dagger\xi_n-[1-(-1)^n]/2$.
Here $x=1/(g^2 a^2)$ and $\mu=2 m_{\rm lat}/ (g^2 a)$, with $a$ the lattice spacing and $m_{\rm lat}$ the bare lattice mass.
The physical Hamiltonian is $H_{\rm phys}=H/(2ax)$. The lattice mass is related to the continuum fermion mass by
$m/g=m_{\rm lat}/g+1/(8\sqrt{x})$~\cite{Dempsey2022,Angelides2023}. In our tensor-network calculations, we use the spin-formulation Hamiltonian, obtained by a
Jordan--Wigner transformation~\cite{Banuls2013} of Eq.~\eqref{eq:H}.

Confinement restricts the physical excitations to neutral mesonic states in this model.
The lightest meson is the vector meson\footnote{In $1+1$
dimensions, this state is also referred to as the pseudoscalar
meson~\cite{Barata2025}.} with charge-conjugation parity $C=-1$,
followed by the heavier scalar meson with $C=+1$~\cite{Hamer1997,Byrnes2002,Banuls2013}.
In the strong-coupling regime, $m/g\ll1$, the scalar is a weakly bound
state of two vector mesons~\cite{Coleman1976}.
At weak coupling, $m/g\gg1$, the scalar is well described by a
fermion--antifermion bound state~\cite{Coleman1976,Bergknoff1977}.

\vspace{0.5cm}
\prlsection{State preparation}
The scalar meson at rest lies above vector-meson states with low kinetic
energy~\cite{Banuls2013}. In a sequential variational matrix product states~(MPS) calculation, reaching the
scalar meson requires first resolving these lower-lying states and enforcing orthogonality to them. 
We instead adapt the QSE construction of Ref.~\cite{Chai2025} to target the scalar meson directly by using selected operators acting on an MPS vacuum~\cite{White1992,Schollwock2011}. By varying this set of operators, we can also test which ones are important for reconstructing the meson.

The $C=-1$ vector meson is dominated by fermion-antifermion operators~\cite{Harada1995}, whereas the scalar meson has a more complex composition. To describe the scalar meson across the mass range, we use three operator classes: two-fermion (2f), compact four-fermion (4f), and meson--meson (MM) operators.

Reflections at the open boundaries produce standing-wave modes rather than plane waves. We therefore sum over the allowed meson positions with sine weights,
\begin{equation}
 w_q(n_{\rm cm})=\sin[k_q(n_{\rm cm}+1)],
 \qquad k_q=\frac{q\pi}{N+1}.
 \label{eq:qse-envelopes}
\end{equation}
Here $n_{\rm cm}$ is the center-of-mass of the meson in lattice units,
and $q=1,2,\ldots,N$ labels the standing-wave modes.

The 2f and compact 4f operators are
\begin{equation}   
\begin{aligned}
O_{(2f,s,d)}(q,c)
&=\sum_{i\in s}w_q(n_{\rm cm})
\xi_i^\dagger\xi_j+c\times(\mathrm{C.c.}),\\
O_{(4f,s,\bm d)}(q,c)
&=\sum_{i\in s}w_q(n_{\rm cm})
\xi_i^\dagger\xi_j\xi_k^\dagger\xi_l+c\times(\mathrm{C.c.}).
\end{aligned}
\label{eq:qse-two-four-fermion}
\end{equation}
The sum runs over reference sites $i$ on the even ($s=e$) or odd
($s=o$) sublattice. For a 2f operator, the second site is $j=i+d$,
where $d$ is a fixed displacement. For a compact 4f operator,
the other three sites are $j=i+d_1$, $k=i+d_2$, and $l=i+d_3$,
with fixed displacements $\bm d=(d_1,d_2,d_3)$.
The centers are $n_{\rm cm}=(i+j)/2$ for 2f, and $(i+j+k+l)/4$ for 4f. Compact 4f operators restrict all four sites to a short interval. The shorthand $\mathrm{C.c.}$ denotes the charge-conjugate partner, added to target the desired $C$ parity, as detailed in Appendix~\ref{app:qse}. For the scalar meson, we use $q=1$ and $c=+1$. With staggered fermions and OBC, $c$ labels an approximate $C$ parity~\cite{Banuls2013}. Finite-size effects break this symmetry and decrease as $1/N$, as in Appendix~\ref{app:approximate-c}.

The MM operators are products of two dressed vector-meson operators,
\begin{align}
    O_{(\mathrm{MM},q_1,q_2)}&=b^\dagger_{q_1}b^\dagger_{q_2}.
    \label{eq:mmoperator}
\end{align}
The internal coefficients of the vector-meson operator $b_q^\dagger$ are determined from the lowest state in a 2f-only QSE calculation with $q=1, c=-1$. We use the same coefficients for all modes $q$. These products describe spatially extended two-meson configurations. We include selected mode pairs with $1\leq q_1\leq q_2\leq3$ to describe their relative motion.

The QSE scalar meson state combines  all three operator classes:
\begin{equation}
    |S_{\rm full}\rangle=\sum_{I\in\mathcal I_{\rm full}}\alpha_I O_I|\Omega\rangle,
\qquad \langle S_{\rm full} | S_{\rm full} \rangle=1,
\end{equation}
Here $I$ is a collective index: $I=(2f,s,d), (4f,s,\boldsymbol d)$, or $(\mathrm{MM},q_1,q_2)$. The set $\mathcal I_{\rm full}$ contains all selected indices from these three operator classes. We determine coefficients $\alpha_I$ by solving a generalized eigenvalue problem in this subspace, as detailed in Appendix~\ref{app:qse}. The scalar meson mass is given by 
\begin{equation}
    \frac{M_S}{g}=\frac{\langle S_{\rm full} |H| S_{\rm full} \rangle-E_{\rm vac}}{2\sqrt{x}},
\end{equation}
where $E_{\rm vac}$ is the vacuum energy.

At $N=200$, $x=50$, and various fermion mass $m_{\rm lat}/g\in\{-1/(8\sqrt{x}),0.125,0.25,0.5,1,5\}$,
we benchmark the QSE scalar meson states against independent variational MPS calculations~\cite{Banuls2013}.
Across all six mass points, the QSE--MPS state fidelities range from $0.9719$ to $0.9945$, and the masses agree within $1.35\%$. These benchmarks support the accuracy of the QSE scalar meson states preparation.
Furthermore, over this mass range, the scalar meson moves from the 11th to the 31st state in the MPS spectrum, counting the vacuum as the first. More detailed comparisons can be found in Table~\ref{tab:dmrgfidelity}.

\vspace{0.5cm}
\prlsection{Scalar composition}
QSE provides information beyond the energy spectrum by identifying which operator classes are needed to represent a particle. Here we examine how the roles of 2f, compact 4f, and MM molecular operators in the scalar meson change with the parameter $m_{\mathrm{lat}} / g$.

Adding compact 4f and then MM operators to the 2f basis gives the nested spaces $A\in\{2f,\,2f+4f,\,2f+4f+\mathrm{MM}\}$.
We solve QSE independently in each space to obtain a normalized scalar state $\ket{S_A}$. Here $\mathcal I_A$ represents the set of operator indices in subspace $A$. The three classes have index sets $\mathcal I_{2f}=\{(2f,s,d)\}$, $\mathcal I_{4f}=\{(4f,s,\bm d)\}$, and $\mathcal I_{\mathrm{MM}}=\{(\mathrm{MM},q_1,q_2)\}$. Thus, $\mathcal I_{2f+4f}=\mathcal I_{2f}\cup\mathcal I_{4f}$ and $\mathcal I_{\mathrm{full}}=\mathcal I_{2f}\cup\mathcal I_{4f} \cup\mathcal I_{\mathrm{MM}}$.
The state obtained in the operator subspace $\mathcal I_A$ and its fidelity to the full QSE state are
\begin{align}
 \ket{S_A}=\sum_{I\in\mathcal I_A}\alpha_I O_I\ket{\Omega}, \ F_A =|\langle S_A|S_{\rm full}\rangle|^2\notag,
 \label{eq:subsetfidelity}
\end{align}
Using the full QSE state as a reference, $F_A$ measures how well the operators indexed by $\mathcal I_A$ represent the scalar meson. The gain in $F_A$ from adding an operator class indicates the importance of that class.
We define the gains from adding compact 4f and then MM operators as $\Delta F_{4f}=F_{2f+4f}-F_{2f}$ and $\Delta F_{\mathrm{MM}}=1-F_{2f+4f}$, respectively, since $F_{\mathrm{full}}=1$.

\begin{figure}[!t]
 \centering
 \includegraphics[width=\linewidth]{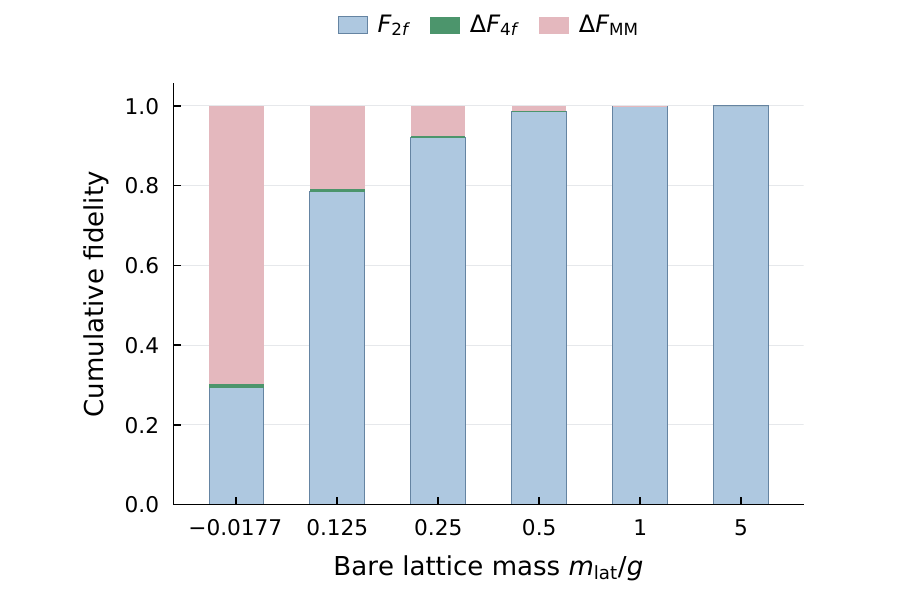}
 \caption{\textbf{Sequential fidelity gains for the scalar meson.} At $N=200$ and $x=50$, blue shows the 2f baseline $F_{2f}$. Green shows the gain $\Delta F_{4f}=F_{2f+4f}-F_{2f}$ from adding compact 4f operators, and rose shows the gain $\Delta F_{\mathrm{MM}}=1-F_{2f+4f}$ from adding MM operators. Each bar sums to $1$ because the full QSE state is the reference. The stacked segments show the 2f fidelity and the gains from adding compact 4f and the MM operators, respectively.}
 \label{fig:subspaces}
\end{figure}

As show in Fig.~\ref{fig:subspaces}, at the improved massless point $m_{\rm lat}/g = -1/(8\sqrt{x}) \approx-0.0177$,  the 2f-only basis give $F_{2f}=0.293$. Adding compact 4f operators gives only $\Delta F_{4f}=0.009$, whereas subsequently adding MM operators gives $\Delta F_{\mathrm{MM}}=0.698$.
The same contrast persists at $m_{\rm lat}/g=0.125$, although the MM gain is smaller. Within the tested basis, spatially extended MM operators are therefore important for reconstructing the light-mass scalar meson, while compact 4f operators provide little improvement. This agrees with the bosonized picture at small fermion masses, where the scalar meson is a weakly bound state of two vector mesons~\cite{Coleman1976,Adam1996}.

As the fermion mass increases, $\Delta F_{\mathrm{MM}}$ decreases. At $m_{\rm lat}/g=1$ and $5$, $F_{2f}$ reaches $0.9987$ and $0.999987$, respectively. The 2f basis is thus sufficient to represent the scalar meson state accurately at large mass, with only small gains $\Delta F_{4f}$ and $\Delta F_{\mathrm{MM}}$ from the additional operators.

\vspace{0.5cm}
\prlsection{Scalar charge and mass response}
We next examine how the operator content identified above affects the meson's scalar charge, which measures its coupling to the scalar fermion current $\bar\psi\psi$~\cite{Gupta2021}. The scalar charge is also related to hadron sigma
terms~\cite{Gupta2021,Holl2006}. For normalized states, its continuum definition and staggered lattice form are
\begin{equation}
\begin{aligned}
 g_S&=\int\dd z\,
 \bigl[\langle\bar\psi(z)\psi(z)\rangle_{S}
       -\langle\bar\psi(z)\psi(z)\rangle_{\Omega}\bigr]\\
 &\simeq\sum_{n=0}^{N-1}(-1)^n
 \bigl(\langle\xi_n^\dagger\xi_n\rangle_{S}
       -\langle\xi_n^\dagger\xi_n\rangle_{\Omega}\bigr).
\end{aligned}
 \label{eq:scalar-charge}
\end{equation}

For exact scalar meson eigenstates, the Feynman--Hellmann theorem relates the scalar charge to the variation of the meson mass with the fermion mass~\cite{Hellmann1937,Feynman1939,GasserZepeda1980}:
\begin{equation}
 \frac{\partial M_S}{\partial m_{\rm lat}} = g_S.
 \label{eq:scalar-charge-fh}
\end{equation}
We evaluate $g_S$ directly using the QSE and MPS scalar meson states as in Eq.~\eqref{eq:scalar-charge}, this gives an estimate of the meson's mass response.

\begin{figure}[t]
 \centering
 \includegraphics[width=\linewidth]{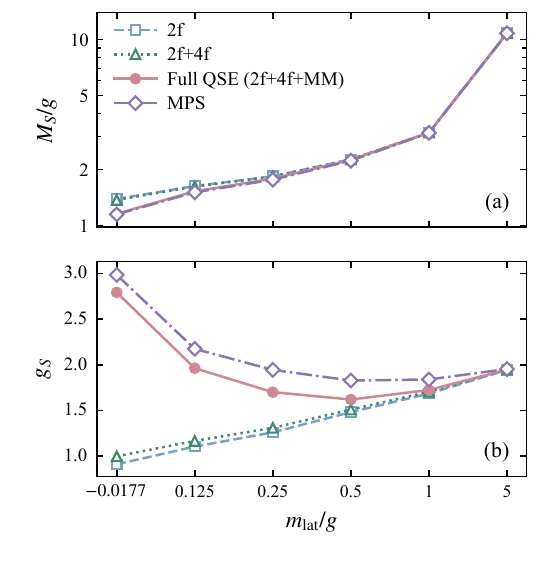}
 \caption{\textbf{Scalar meson mass and scalar charge.}
 (a) Scalar meson mass $M_S/g$ and (b) scalar charge
 $g_S$ at $N=200$ and $x=50$.
 Blue open squares, green open triangles, and rose filled circles
 represent the independently optimized 2f, 2f+4f, and full
 $2f+4f+\mathrm{MM}$ QSE states, respectively.
 Purple open diamonds represent results for the MPS states obtained by the two-site variational optimization algorithm. The six fermion masses are shown at equally spaced positions for readability.
 Panel (a) uses a logarithmic vertical axis, where lines are added to guide the eye.
 All QSE and MPS states have bond dimensions up to 300 and 200, respectively.}
 \label{fig:scalar-charge}
\end{figure}
Figure~\ref{fig:scalar-charge} compares
the meson mass and scalar charge as the QSE operator basis is enlarged. We use the full QSE results as the reference.
At large fermion mass, the 2f-only basis produce both $M_S/g$ and $g_S$ closely to the full QSE value. As the fermion mass decreasing, the deviation from the full QSE appears, especially for $g_S$. Specifically, at $m_{\rm lat}/g = 0.25$, the 2f meson mass $M_S/g$ is still close to the full QSE value, while 2f scalar charge $g_S$ is significantly lower.
At light fermion mass, adding compact four-fermion operators changes the meson mass little and gives a modest increase in \(g_S\). Meson–meson operators account for most of the remaining increase toward the full QSE scalar charge. In a short summary, $g_S$ is more sensitive to the operator content
than the meson mass, particularly to the addition of MM operators.

The independent MPS results provide a separate check of the full QSE calculation. The meson masses agree well across the fermion masses we considered, but the full QSE scalar charges are noticeably lower than the MPS values. At $m_{\rm lat}/g \approx-0.0177$, $g_S$ is still about $6.4\%$ below the MPS value despite a QSE--MPS state fidelity of $0.984$. Agreement in mass and even a high state fidelity therefore do not guarantee equally accurate scalar charges. These differences indicate the need for further convergence checks, including error accumulating effect in the excited-state calculation with MPS, testing more MM modes in QSE and larger system sizes. We leave these systematic studies to future work and focus here on the physical trends.

\vspace{0.5cm}
\prlsection{PDF asymmetry and scalar composition}
Having examined the scalar charge, we now test how the meson composition identified by QSE is reflected in the PDF
$f_S(\xi)$, a key quantity for describing hadron structure~\cite{Feynman1969,BjorkenPaschos1969}.
In the parton picture, $f_S(\xi)$ describes the probability density of finding a fermion carrying a momentum fraction $\xi=k^+/P^+$ of the meson, where $k^+$ and $P^+$ are the fermion and meson light-cone momenta, respectively.

The PDF can be defined by a Fourier transform of a light-cone correlator:
\begin{align}
 \mathcal C_S(z^-)={}&\bra{S(P)}\bar\psi(z^-)\Wline(z^-,0)
 \gamma^+\psi(0)\ket{S(P)}_c,
 \label{eq:Cpdf}\\
 f_S(\xi)&=\int_{-\infty}^{\infty}\frac{\dd z^-}{4\pi}\,
 e^{-\ii\xi P^+z^-}\mathcal C_S(z^-)\notag\\
 &\simeq\frac{M_S aN}{4\pi}
 \sum_{l=0}^{L-1} e^{\ii\xi M_S 2a(l-l_s)}C_S(l, l_s).
 \label{eq:pdf}
\end{align}
The lattice approximation follows Ref.~\cite{Banuls2026}. The state $\ket{S(P)}$ is the scalar meson with momentum $P$. We use
the lowest mode $\ket{S}$ in the full QSE calculation, corresponding to $P=0$ up to finite volume effects.
The Wilson line $\Wline$ ensures gauge invariance, and the subscript "$c$" denotes subtraction of the vacuum contribution.
We group each pair of staggered sites $n=(2l,2l+1)$ into a Dirac site,
with the even and odd sites carrying the upper and lower Dirac spinor
components. The resulting $L=N/2$ Dirac sites have lattice spacing $2a$.
$C_S(l,l_s)$ is the lattice correlator between site $l$ and the fixed
source site $l_s$, separated by $z = 2a(l-l_s)$.
Implementation details are given in Appendix~\ref{app:correlators}.

\begin{figure}[!t]
 \centering
 \includegraphics[width=\linewidth]{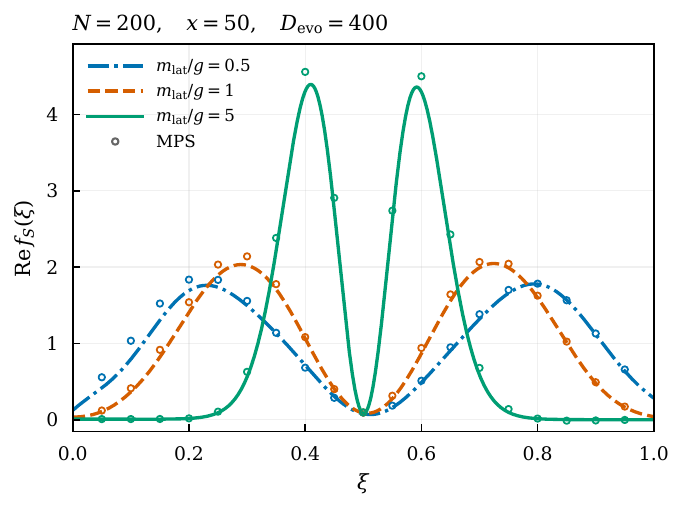}
 \caption{\textbf{Scalar PDFs at heavier masses.}
PDFs at $N=200$, $x=50$, and $D_{\rm evo}=400$ for full QSE and matched
MPS states. The MPS states are obtained using the two-site variational optimization algorithm.
 Real parts of the PDFs from Eq.~\eqref{eq:pdf} are shown for $0\leq\xi\leq1$.
 Dash-dotted, dashed, and solid QSE curves identify masses 0.5, 1, and 5.
 Same-color open circles show the MPS results at selected points.}
 \label{fig:pdfheavy}
\end{figure}

Figure~\ref{fig:pdfheavy} shows the scalar meson PDFs at the heavier fermion masses
$m_{\rm lat}/g\geq0.5$, with two peaks and a minimum close to zero at
$\xi=1/2$. The matched MPS results reproduce this main shape.
At these masses, the 2f subspace suffices to
represent the scalar meson (Fig.~\ref{fig:subspaces}). 
In the light-front two-body sector, containing one fermion and one
antifermion, the scalar meson wave function satisfies
$\psi_2(\xi,1-\xi)=-\psi_2(1-\xi,\xi)$~\cite{Harada1995}.
Therefore, the PDF is proportional to $|\psi_2|^2$, symmetric about
$\xi=1/2$, and zero at that point.
The observed shape is consistent with this two-body picture. However, QSE
bilinears acting on the interacting vacuum need not produce a pure
two-body Fock state. Higher Fock contributions and finite-volume,
lattice-spacing, and Trotter effects may give a nonzero $f_S(1/2)$.
We leave a systematic study of these effects to future work.

As the fermion mass increases, the two peaks become narrower and move
closer to $\xi=1/2$. A similar trend is seen in the vector meson PDFs
of Ref.~\cite{Banuls2026}, where the peak at $\xi=1/2$ becomes narrower
at larger fermion masses.

At light fermion masses $m_{\rm lat}/g\leq0.125$, MM molecular operators are important,
so the PDF need not retain the symmetry about $\xi=1/2$ found in the two-body sector.
However, the correlators decay more slowly at light masses, and the PDFs have not converged at $N=200$ (Appendix~\ref{app:correlators}). Reconstructing the full PDFs at these light masses requires larger systems. To evaluate whether the PDF is symmetric about $\xi=1/2$ in current setup, we therefore use a finite-distance correlator probe that does not require full PDF reconstruction. Specifically, we probe departures from the two-body reflection symmetry using
\begin{equation}
 \begin{aligned}
 B_1&=\int_0^1\dd\xi\,f_S(\xi)\sin(2\pi\xi)\\
 &=\frac{N}{16\ii}
 \bigl[C_S(l_-^*,l_s)-C_S(l_+^*,l_s)\bigr].
 \end{aligned}
 \label{eq:B1}
\end{equation}
A symmetric PDF has $B_1=0$ since the $\sin$ function is antisymmetric at $\xi=1/2$. A positive $B_1$ indicates that fermions in the meson prefer taking momentum fractions $\xi<1/2$, weighted by the sine function.
The correlator estimate follows from the Fourier transformation derived in Appendix~\ref{app:correlators}, with $l_\pm^*=l_s\pm\pi/(aM_S)$.
Thus only the correlator near $l_\pm^*$ is needed, avoiding the long-distance correlators that have large finite volume effects. Since the positions $l_\pm^*$ are
generally not integers, we use four-point cubic interpolation of the correlator around $l_\pm^*$.

\begin{figure}[!t]
 \centering
 \includegraphics[width=\linewidth]{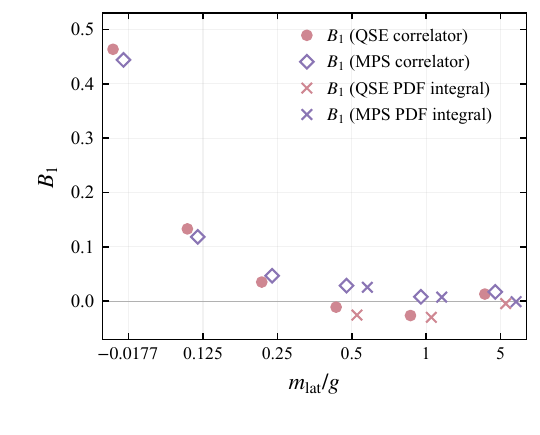}
 \caption{\textbf{PDF asymmetry across the scalar mass scan.}
 At $N=200$, $x=50$, and $D_{\rm evo}=400$, rose filled circles
 and purple open diamonds show $B_1$
 from the correlator in Eq.~\eqref{eq:B1} for full QSE and MPS states obtained using two-site variational optimization, respectively.
 Rose and purple crosses show the corresponding QSE and MPS PDF-integral
 results at $m_{\rm lat}/g=0.5,1,5$, as in Eq.~\eqref{eq:B1FT}.
 Only real parts are displayed.
 The mass points are equally spaced, with small marker offsets for
 readability.}
 \label{fig:B1}
\end{figure}

We first check Eq.~\eqref{eq:B1} against the heavier-mass PDFs with $m_{\rm lat}/g\geq0.5$,
where the full PDFs are reconstructed as in Fig.~\ref{fig:pdfheavy}. This allows a cross-check of $B_1$ from PDF integration and direct correlator evaluation in Eq.~\eqref{eq:B1}. Both methods
give small values, consistent with the approximately symmetric shapes
in Fig.~\ref{fig:pdfheavy}. Their differences can arise from the finite correlator range used to reconstruct the PDF and from correlator interpolation.

At light masses, we instead use the correlator estimate without relying
on an unconverged full PDF. Figure~\ref{fig:B1} shows an increasingly positive
$B_1$ as the fermion mass decreases, indicating that fermions have a higher probability of carrying smaller momentum fractions $\xi<0.5$. At the improved massless point, the QSE and independently
optimized MPS states give 0.4634 and 0.4439, respectively.
Both approaches show the same trend, the increasing PDF asymmetry accompanies the growing importance of MM operators. This trend is consistent with a two-meson picture, in which more fermionic constituents share the momentum of the meson.

\vspace{0.5cm}
\prlsection{Conclusion and outlook}
We used QSE to study the internal structure of the scalar meson in the massive Schwinger model. At large fermion masses, fermion bilinears suffice to reconstruct the scalar meson. While as the fermion mass is decreasing, the four-fermion content becomes essential. Specifically, the compact four-fermion operators give little improvement, while meson–meson operators are essential for the construction of the scalar meson.

For the observables, the scalar charge can change substantially as the operator basis is enlarged, even when the meson mass is stable. As the fermion mass is  decreasing and meson–meson operators become more important, the PDF shows an asymmetry of $\xi=0.5$. This asymmetry can be evaluated by a suitable correlator at finite separation, even in cases when the full PDF is difficult to reconstruct. Together, these results show how QSE can quantify the importance of different operator classes in reconstructing a composite particle and connect its internal structure to physical observables.

Quantitative control requires further convergence tests of both the prepared states and the calculation of PDF, including larger operator basis and studies of finite-volume and lattice-spacing effects. In the future, we also plan to extend this operator-based approach to non-Abelian gauge theories and higher dimensions using tensor networks or quantum simulation. This will require gauge-invariant operator bases suited to the states of interest. For example, including gluonic operators such as traced plaquettes would allow us to test how much they improve the reconstruction of composite states.

\vspace{0.5cm}
\begin{acknowledgments}
\prlsection{Acknowledgments}
We thank Fernanda Steffens, Aniket Sen, Lukas Mansour for discussions about the PDF in Schwinger model. Yahui Chai also thanks Yan Lyu for discussions on the sigma term
and Hongxi Xing for discussions on parton distributions. All tensor-network calculations were performed using the ITensor library~\cite{ITensor}, with U(1) symmetry implemented. Computations were performed on the DESY Zeuthen computing facilities.  This work was supported by the European Union's Horizon Europe Framework Programme under the ERA Chair scheme (Grant Agreement No.~101087126) and by the RIKEN TRIP initiative (RIKEN Quantum). Codex and Claude were used to assist with code development, figure preparation, and partial draft polishing. All AI-assisted materials were reviewed and approved by the authors, who take full responsibility for the results presented in this work.
\end{acknowledgments}

\bibliography{refs}

\setcounter{secnumdepth}{2}
\appendix

\section{Eighth-order decompositions}\label{app:highorder}
This appendix section introduces the eighth-order decompositions of the time-evolution operator. We include a static charge in
the dimensionless Hamiltonian of Eq.~\eqref{eq:H}~\cite{Banuls2026}. This Hamiltonian is split into three parts,
\begin{equation}\label{eq:trotter-split}
  H=H_A+H_B+H_z,
\end{equation}
with
\begin{equation}
  H_A=\sum_{n\ \mathrm{even}}h_n,\qquad
  H_B=\sum_{n\ \mathrm{odd}}h_n.
\end{equation}
Here,
\begin{equation}
    h_n=-\ii x\bigl(\xi_n^\dagger\xi_{n+1}-\mathrm{h.c.}\bigr)
\end{equation}
for $0\le n\le N-2$, and
\begin{equation}\label{eq:trotter-Hz}
  H_z=\sum_{n=0}^{N-2}
          L_n^2+\mu\sum_{n=0}^{N-1}(-1)^n\xi_n^\dagger\xi_n,
\end{equation}
with $L_n=\sum_{j=0}^{n}(Q_j+q^{\mathrm{st}}_j)$,
$Q_j=\xi_j^\dagger\xi_j-[1-(-1)^j]/2$, and $L_{-1}=0$.
The static charge $q^{\mathrm{st}}_j$ is introduced to implement the Wilson line. It remains at site $n_{\mathrm{st}}$ during each time interval, with
$q^{\mathrm{st}}_j=\delta_{j,n_{\mathrm{st}}}$. The position $n_{\mathrm{st}}$ is updated between intervals to follow the Wilson-line path. For more details, we refer the reader to Ref.~\cite{Banuls2026}.

We define the layers $\mathcal A(\tau)=e^{\mathrm i\tau H_A}$, $\mathcal B(\tau)=e^{\mathrm i\tau H_B}$, and $\mathcal Z(\tau)=e^{\mathrm i\tau H_z}$. For real $\tau$, all these layers are unitary. We refer to $\mathcal A$ and $\mathcal B$ as hopping layers. They are products of two-site unitary gates on the links $(0,1),(2,3),\dots$ and $(1,2),(3,4),\dots$, respectively. We refer to $\mathcal Z$ as the diagonal layer. It acts on all sites and implements evolution under the mass and electric terms.

For each time interval, we divide the duration $2a$ into $N_{\tau}$ time steps.
Since $H_{\mathrm{phys}}=H/(2ax)$, the exact evolution operator for one step is
\begin{equation}\label{eq:trotter-step}
  e^{\mathrm{i}H_{\mathrm{phys}}\,\eta\,2a/N_\tau}
  =e^{\mathrm i\eta\tau H},
  \qquad \tau=\frac{1}{xN_\tau},
\end{equation}
where $\eta=\operatorname{sgn}(l-l_s)$ denotes the evolution direction,
with $l$ and $l_s$ the sink and source Dirac-site labels in Eq.~\eqref{eq:pdf}.
In our simulation, we set $N_{\tau}$ as $4$.
The symmetric second-order Trotter approximations
to $e^{\mathrm i\tau H}$~\cite{Suzuki1990},
\begin{equation}
  K_B(\tau)=\mathcal A(\tfrac{\tau}{2})\,
             \mathcal Z(\tfrac{\tau}{2})\,
             \mathcal B(\tau)\,
             \mathcal Z(\tfrac{\tau}{2})\,
             \mathcal A(\tfrac{\tau}{2}),\label{eq:trotter-KB}
\end{equation}
serve as the kernels for the eighth-order decompositions introduced below. After adjacent $\mathcal A$ layers are merged, an $m$-stage composition contains $2m+1$ hopping layers with $K_B$. Only hopping layers require singular-value decompositions (Appendix~\ref{app:blockphase}).

Kahan and Li~\cite{KahanLi1997} proposed a 17-stage
eighth-order composition. Using the kernel $K_B$, we have
\begin{equation}\label{eq:trotter-S8}
  \mathcal S_8(\tau)=K_B(\gamma_{17}\tau)\cdots K_B(\gamma_2\tau)\,K_B(\gamma_1\tau).
\end{equation}
We impose $\gamma_j=\gamma_{18-j}$, leaving nine independent weights. The composition is then palindromic and satisfies $\mathcal S_8(-\tau)=\mathcal S_8(\tau)^{-1}$. Its logarithm therefore contains only odd powers of $\tau$.
The stage weights satisfy the normalization condition $\sum_{j=1}^{17}\gamma_j=1$, which ensures a total step of $\tau$. Besides, they satisfy seven conditions that cancel all independent error terms at orders $\tau^3$, $\tau^5$, and $\tau^7$ in the logarithmic expansion~\cite{KahanLi1997}. These conditions give $\log\mathcal S_8(\tau)=\mathrm i\tau H+\mathcal{O}(\tau^9)$. Equivalently, $\mathcal S_8(\tau)=e^{\mathrm i\tau H}+\mathcal{O}(\tau^9)$.

We use the coefficient set \texttt{s17odr8a}~\cite{LiKahanNetlib}. The nine independent weights are listed in Table~\ref{tab:trotter-weights}. Our code uses 26-decimal values that round to these (see also Ref.~\cite{Lubich2008}, Sec.~III.3.3). They give the same double-precision numbers as the 20-digit values listed in the table.
\begin{table}[t]
\caption{Weights of the Kahan--Li composition
\texttt{s17odr8a}~\cite{LiKahanNetlib}, with $\gamma_j=\gamma_{18-j}$. Values are given to 20 decimal places.}
\label{tab:trotter-weights}
\centering\small
\begin{tabular}{@{}lr@{}}
\toprule
$j$ & $\gamma_j$ \\ \midrule
1,\ 17 & $0.13020248308889008088$ \\
2,\ 16 & $0.56116298177510838456$ \\
3,\ 15 & $-0.38947496264484728641$ \\
4,\ 14 & $0.15884190655515560090$ \\
5,\ 13 & $-0.39590389413323757734$ \\
6,\ 12 & $0.18453964097831570709$ \\
7,\ 11 & $0.25837438768632204729$ \\
8,\ 10 & $0.29501172360931029887$ \\
9 & $-0.60550853383003451170$ \\
\bottomrule
\end{tabular}
\end{table}
For each interval of duration $2a$, we approximate $e^{\mathrm iH_{\mathrm{phys}}\eta\,2a}=(e^{\mathrm i\eta\tau H})^{N_\tau}$ by $[\mathcal S_4(\eta\tau)]^{N_\tau}$ or $[\mathcal S_8(\eta\tau)]^{N_\tau}$. 

\section{The electric layer as block phases}\label{app:blockphase}
In Appendix~\ref{app:highorder}, each diagonal layer applied to the MPS is $\mathcal Z(t)=e^{\mathrm itH_z}$ with real $t$, equal to $\eta\tau$ times the coefficient of the layer (for example, $t=\eta\gamma_j\tau/2$ in $K_B$), and
$H_z$ is defined in Eq.~(\ref{eq:trotter-Hz}). We use U(1) symmetry in our calculations. Each symmetry sector of the virtual bond after site $n$ corresponds to a definite value of the accumulated fermion number $\sum_{j=0}^{n}\xi_j^\dagger\xi_j$. Together with the staggered background and static-charge contributions, this fixes the electric field $L_n$ on the corresponding link. Dempsey et al.~\cite{Dempsey2026} use this representation of the electric field through virtual-bond charge operators in their construction. Here we use this representation to apply the diagonal layer as block phases during real-time evolution.

To determine the phase of each block, we label each symmetry
sector of the virtual bond after site $n$ by $b_n$, and denote its
accumulated fermion number by $M_n(b_n)$. Gauss's law gives
\begin{equation}
 L_n(b_n)=M_n(b_n)-\sum_{j=0}^{n}\frac{1-(-1)^j}{2}+\sum_{j=0}^{n}q_j^{\mathrm{st}}.
 \label{eq:electric-bond-charge}
\end{equation}
This determines the electric contribution to the block phase. The
mass contribution is fixed by the local occupation $o_n=0,1$, the
eigenvalue of $\xi_n^\dagger\xi_n$. Together, these contributions
determine the phase applied to each tensor block. Let
$\mathsf A^{[n]o_n}_{b_{n-1}b_n}$ denote a tensor block at site $n$.
Each block is updated as
\begin{equation}\label{eq:electric-block-phase}
\begin{aligned}
  \mathsf A^{[n]o_n}_{b_{n-1}b_n}
  \mapsto{}&
  e^{\mathrm it\mu(-1)^n o_n}
  \mathsf A^{[n]o_n}_{b_{n-1}b_n}
  \\
  &\times
  \begin{cases}
    e^{\mathrm it
       L_n(b_n)^2},
      & 0\le n\le N-2,\\[1ex]
    1, & n=N-1.
  \end{cases}
\end{aligned}
\end{equation}
All entries within a block are multiplied by the same phase.
The last tensor receives only the mass phase. The update is exact for the input MPS and requires no additional cutoff on the electric flux and no singular-value decomposition.

\section{QSE construction and validation}
\label{app:qse}

\subsection{Charge conjugation and operator basis}
\label{app:approximate-c}

We follow Ref.~\cite{Chai2025}. All operators conserve total charge.
For staggered fermions, charge conjugation combines particle--hole
exchange with a one-site translation~\cite{Banuls2013}. We use
\begin{equation}
 C\xi_n^\dagger C^{-1}=\xi_{n+1},\qquad
 C\xi_n C^{-1}=\xi_{n+1}^\dagger.
 \label{eq:qse-c-staggered}
\end{equation}
For distinct sites,
fermion anticommutation gives
\begin{equation}
 C(\xi_i^\dagger\xi_{i+d})C^{-1}
 =-\xi_{i+d+1}^\dagger\xi_{i+1},\qquad d>0.
 \label{eq:qse-c-bilinear}
\end{equation}
\emph{2f operators under PBC.} Let $s=e,o$ label the even and odd
sublattices, respectively, with $i\in s$ and $\bar s$ the opposite
sublattice. For a fixed separation $d>0$, define
\begin{align}
 O^{\rm PBC}_{s,d}(k,c)={}&
 \sum_{i\in s}e^{\ii k(i+d/2+1)}\xi_i^\dagger\xi_{i+d}\notag\\
 &-c\sum_{i\in\bar s}e^{\ii k(i+d/2+1)}\xi_{i+d}^\dagger\xi_i,
 \label{eq:qse-twofermion-pbc}
\end{align}
where $k=2\pi r/N$ for integer $r$, and $c=\pm1$.
Using Eq.~\eqref{eq:qse-c-bilinear}, we obtain
\begin{equation}
 CO^{\rm PBC}_{s,d}(k,c)C^{-1}
 =c\,e^{-\ii k}O^{\rm PBC}_{s,d}(k,c).
 \label{eq:qse-c-pbc}
\end{equation}
The phase comes from the one-site translation; at $k=0$ the
operator has $C$ parity $c$.

\emph{2f operators under OBC.} We combine plane-wave envelopes
with opposite momenta into the standing-wave envelope $w_q$ in
Eq.~\eqref{eq:qse-envelopes}. The full operator is
\begin{align}
 O_{(2f,s,d)}(q,c)={}&
 \sum_{i\in s}w_q\!\left(i+\frac d2\right)
 \xi_i^\dagger\xi_{i+d}\notag\\
 &-c\sum_{i\in\bar s}w_q\!\left(i+\frac d2\right)
 \xi_{i+d}^\dagger\xi_i.
 \label{eq:qse-twofermion-expanded}
\end{align}
Both sums advance by two sites and keep all sites inside the
chain, $0\leq i\leq N-1-d$. The weight is evaluated at the center
$n_{\rm cm}=i+d/2$ of each bilinear. The QSE uses $q=1$,
$d=0$, and odd $d\leq L_{2f}=40$.

For $d>0$, write this sine-weighted operator $O_{(2f,s,d)}$ as $O_c^{\sin}$ and
let $O_c^{\cos}$ denote the same sums with cosine weights.
The one-site translation in $C$ gives
\begin{align}
 CO_c^{\sin}C^{-1}
 ={}&c\left(\cos k_q\,O_c^{\sin}
       -\sin k_q\,O_c^{\cos}\right)\notag\\
 &+\Delta_{\partial},
 \label{eq:qse-c-sine-mixing}
\end{align}

where $\Delta_{\partial}$ contains the boundary correction terms from the one-site shift. At fixed $q$ and $d$, only a fixed number
of bilinears contribute to $\Delta_{\partial}$. Their centers $n_{\rm cm}$ close to the boundary and with the weight $w_q = \sin\left( (n_{\rm cm}+1)\pi/(N+1) \right) \sim O(1/N)$. Thus $\|\Delta_{\partial}\|=O(1/N)$,
and the $O_c^{\cos}$ coefficient $\sin k_q$ is also $O(1/N)$.The approximate $C$ parity therefore improves with increasing volume.

\emph{Compact 4f operators under OBC.} Here $s=e,o$ labels the sublattice of
the reference site $i$, and $\bm d=(d_1,d_2,d_3)$ specifies the
relative positions of the other three sites. The full expression
in Eq.~\eqref{eq:qse-two-four-fermion} is
\begin{align}
 O_{(4f,s,\bm d)}(q,c)={}&
 \sum_{i\in s}w_q(n_{\rm cm})
 \xi_i^\dagger\xi_j\xi_k^\dagger\xi_l\notag\\
 &+c\sum_{i\in\bar s}w_q(n_{\rm cm})
 \xi_j^\dagger\xi_i\xi_l^\dagger\xi_k.
 \label{eq:qse-fourfermion-expanded}
\end{align}
Here $j=i+d_1$, $k=i+d_2$, and $l=i+d_3$ in both sums, with
center $n_{\rm cm}=(i+j+k+l)/4$.
The reference site $i$ need not be leftmost, so the separations
may be negative; all sites must remain inside the chain. We retain four distinct sites with separation at most
$L_{4f}=5$ lattice spacings, ordered as $d_2>0$ and $d_1<d_3$.

The charge conjugation transformation to the compact 4f operators can result in a similar formula as Eq.~\eqref{eq:qse-c-sine-mixing}. Both the coefficient of cosine weighted operator and the norm of the boundary correction are therefore $O(1/N)$.

\emph{MM operators under OBC.}
For $O_{\rm MM}=b_{q_1}^\dagger b_{q_2}^\dagger$, charge conjugation
acts on each factor separately,
\[
 C O_{\rm MM} C^{-1}
 = (C b_{q_1}^\dagger C^{-1})(C b_{q_2}^\dagger C^{-1}).
\]
Using Eq.~\eqref{eq:qse-c-sine-mixing} with $c=-1$, we obtain
\[
 C b_q^\dagger C^{-1}
 = -\bigl(\cos k_q\,b_q^\dagger
          -\sin k_q\,b_{q,\cos}^\dagger\bigr)
   +\Delta_{\partial,q},
 \qquad q=q_1,q_2,
\]
where $b_{q,\cos}^\dagger$ is obtained from $b_q^\dagger$ by
replacing the sine weights with cosine weights.
As discussed above, the coefficient $\sin k_q$ and the
boundary correction norm $\|\Delta_{\partial,q}\|$ are both
$O(1/N)$.So we can write the product as
\begin{align*}
 C O_{\rm MM} C^{-1}
 &= \bigl(\cos k_{q_1}\,b_{q_1}^\dagger + O(1/N)\bigr)
    \bigl(\cos k_{q_2}\,b_{q_2}^\dagger + O(1/N)\bigr)\\
 &= \cos k_{q_1}\cos k_{q_2}\,O_{\rm MM} + O(1/N).
\end{align*}
At fixed $q_1$ and $q_2$, both cosine factors approach unity as
$N$ increases, giving the approximate $C=+1$ assignment.

\subsection{Projected problem and independent checks}

For $\ket{\phi_i}=O_i\ket{\vac}$, the projected matrices are
\begin{align}
 H_{ij}&=\bra{\phi_i}H\ket{\phi_j},\qquad
 S_{ij}=\langle\phi_i|\phi_j\rangle,\notag\\
 Z_{ij}&=\bra{\vac}O_jO_i^\dagger\ket{\vac}.
 \label{eq:qsemat}
\end{align}
The matrix $Z$ penalizes the action of the corresponding annihilation
operator on the vacuum, following Ref.~\cite{Chai2025}. After removing
near-null directions of $S$, we solve
\begin{equation}
 (H+Z)\bm\alpha=\lambda S\bm\alpha,\qquad
 \bm\alpha^\dagger S\bm\alpha=1.
 \label{eq:qseproblem}
\end{equation}
We select the lowest non-vacuum candidate in the scalar-targeted basis.
The eigenvalue $\lambda$ includes the $Z$ penalty; the energy is instead $E=\bra{S}H\ket{S}$, measured on the normalized
reconstructed MPS. The mass is
$M_S/g=(E-E_{\rm vac})/(2\sqrt{x})$.

Table~\ref{tab:dmrgfidelity} compares QSE and independently optimized
MPS states, the state fidelity is
$F_{\rm MPS}=|\langle S_{\rm MPS}|S_{\rm full}\rangle|^2$. The QSE energy standard deviation
\begin{equation}
 \frac{\sigma_E}{g}
 =\frac{\sqrt{\bra{S}H^2\ket{S}-E^2}}{2\sqrt{x}}.
 \label{eq:energyfluctuation}
\end{equation}
The factor $1/(2\sqrt{x})$ converts to units of $g$.
The standard deviation vanishes for an exact eigenstate.

The reference MPS states are obtained by two-site variational
optimization with a maximum bond dimension of 200 and an SVD cutoff
of $10^{-10}$. We use an SVD cutoff of $10^{-8}$ for QSE basis
construction and full-state reconstruction. The QSE reconstruction
uses a maximum bond dimension of 300 at $m_{\rm lat}/g\leq0.5$
and 100 at $m_{\rm lat}/g=1,5$.

\begin{table}[!htbp]
\centering
\caption{Comparison of full QSE and independently optimized MPS
scalar states at $N=200$ and $x=50$. The MPS state index
$n_S^{\rm MPS}$ counts the vacuum as state 1. $F_{\rm MPS}$ is the fidelity between the full QSE state and the independently optimized MPS state.}
\label{tab:dmrgfidelity}
\small
\setlength{\tabcolsep}{2pt}
\begin{tabular}{@{}cccccc@{}}
\hline
$m_{\rm lat}/g$ & $n_S^{\rm MPS}$ & $M_S^{\rm QSE}/g$ & $M_S^{\rm MPS}/g$
 & $F_{\rm MPS}$ & $\sigma_E/g$ \\
\hline
$-1/(8\sqrt{x})$ & $11$ & $1.15753$ & $1.14918$ & $0.9837$ & $0.22796$ \\
$0.125$ & $13$ & $1.52882$ & $1.50983$ & $0.9719$ & $0.21976$ \\
$0.25$ & $14$ & $1.78855$ & $1.76484$ & $0.9733$ & $0.23178$ \\
$0.5$ & $16$ & $2.25248$ & $2.23324$ & $0.9761$ & $0.23020$ \\
$1$ & $19$ & $3.15858$ & $3.14875$ & $0.9945$ & $0.20243$ \\
$5$ & $31$ & $10.81427$ & $10.81373$ & $0.9913$ & $0.08474$ \\
\hline
\end{tabular}
\end{table}
% Updated 2026-10-23; measurements and source hashes: yibin/table1_update_20261023/.
% Reproduction: scripts/analysis/update_manuscript_table1_20261023.jl.
% DMRG: latest yibin/data/int/data_scalar/.../Epsi/psi{k}.h5 snapshots,
% not the earlier ritz_psi{k}.h5 copies; k=11,13,14,16,19,31 in row order.
% Shared vacuum: yibin/yibin_states/.../psi1.h5, checked against DMRG E.h5.
% QSE first four rows: yibin/QSE_D300_states_20261022_161940/.../subset_states.h5.
% QSE last two rows: data/...Dmax100.../QSE/kind1_L2f40_L4f5_pmm1_modes3_2fcm_4fcm_c1/subset_states.h5.
% All QSE rows use psi_2f_4f_pmm; energies and variances are measured on normalized MPS,
% not projected-matrix eigenvalues. Site IDs are aligned without truncation for fidelity.

\section{Light-cone correlators}\label{app:correlators}
We implement the Wilson line as a moving static charge, following
Ref.~\cite{Banuls2026}. The higher-order evolution
schemes are described in Appendix~\ref{app:highorder}, with the
diagonal-layer implementation given in Appendix~\ref{app:blockphase}.
Each Dirac site $l=0,\ldots,L-1$ contains staggered sites
$(2l,2l+1)$ carrying the upper and lower Dirac spinors, with $L=N/2$
and spacing $2a$. 

We fix the source at $l_s=L/2$. Its two staggered sites are
$n_s(l_s)=2l_s+s$, with $s=0,1$. The corresponding sink sites
are $n_r(l)=2l+r$, with $r=0,1$. At fixed $l_s$, we also write $C_S(l,l_s)$ as $C_S(\Delta n)$
or $C_S(\nu)$, where $\Delta n=2(l-l_s)$ is the separation in
staggered-site units, and the Ioffe time $\nu$ is defined as
\[
 \nu=-P^+z^-=aM_S\Delta n=2aM_S(l-l_s).
\]

\begin{figure}[!htbp]
 \centering
 \includegraphics[width=\linewidth]{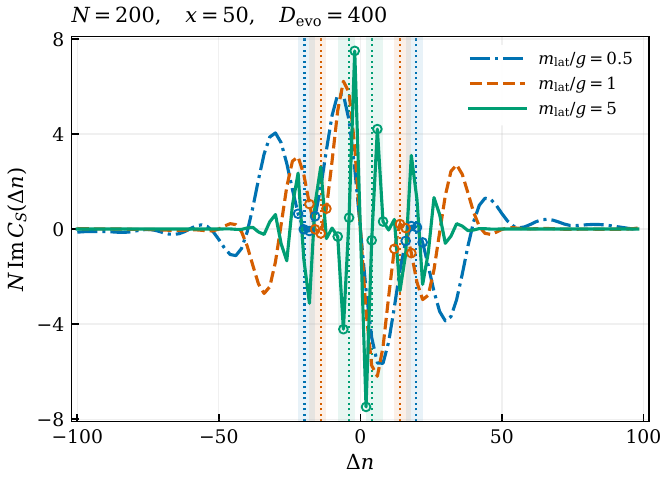}
 \caption{\textbf{Light-cone correlators at heavier masses.}
 Full QSE states at $N=200$, $x=50$, and $D_{\rm evo}=400$.
 We show $N\operatorname{Im}C_S(\Delta n)$, with vacuum subtraction,
 at all measured separations. Dash-dotted, dashed, and solid lines
 identify masses 0.5, 1, and 5.
 Vertical dotted lines mark $\nu=\pm2\pi$. Matching bands span the four
 measured points used for cubic interpolation in each direction.
 Open circles mark these points. PDF at these masses are shown in Fig.~\ref{fig:pdfheavy}.}
 \label{fig:corrheavy}
\end{figure}

\begin{figure}[!htbp]
 \centering
 \includegraphics[width=\linewidth]{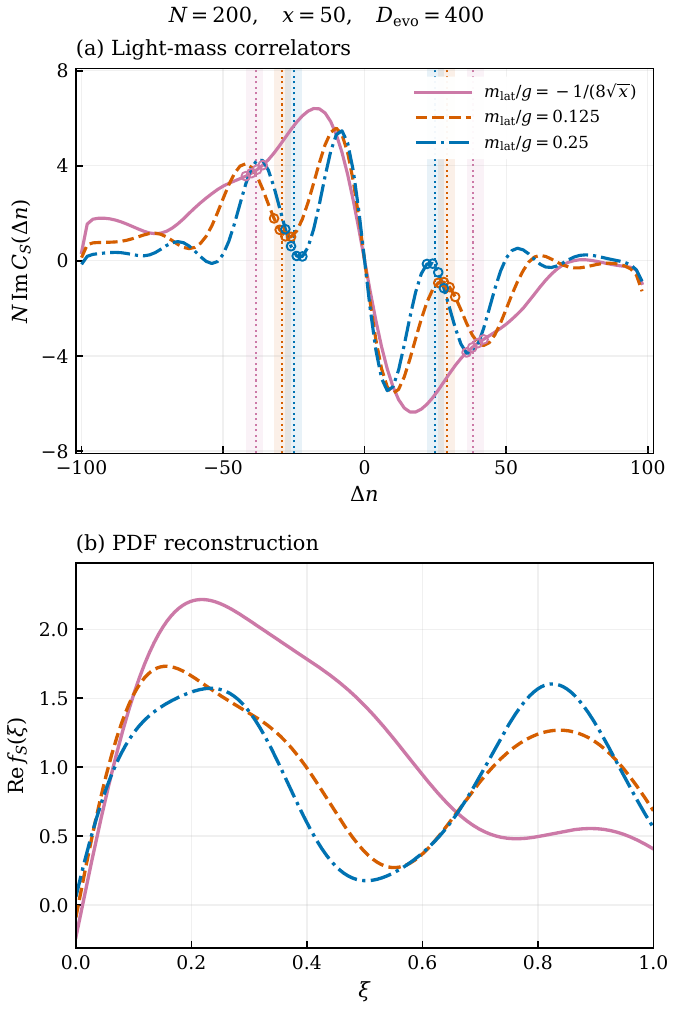}
 \caption{\textbf{Long-distance correlators and finite-range PDF
 reconstructions at lighter masses.}
 The improved massless point and $m_{\rm lat}/g=0.125,0.25$, with the same
 parameters as Fig.~\ref{fig:corrheavy}.
 (a) $N\operatorname{Im}C_S(\Delta n)$, with the same vertical scale and
 interpolation-window conventions as Fig.~\ref{fig:corrheavy}.
 (b) Real parts of the PDFs from all measured correlators, shown for
 $0\leq\xi\leq1$. Colors and line styles match panel (a). The light-mass PDFs are shown only to illustrate the
 current reconstruction.
}
\label{fig:corrlight}
\end{figure}

The PDFs and $B_1$ calculations in Fig.~\ref{fig:pdfheavy} and \ref{fig:B1} uses eighth-order \texttt{kl8qn} evolution,
four substeps per $2a$, and an SVD cutoff of $10^{-14}$.
The evolution bond limit is $D_{\rm evo}=400$. Increasing it from 300 to 400 changes the displayed PDFs
by less than $0.5\%$ of their peak heights and the
correlator-based $B_1$ values by less than $0.12\%$
at all six fermion masses. For each state, we subtract a separately evolved vacuum with matching settings. The correlators for these heavy mass are shown in Figs.~\ref{fig:corrheavy}, whose value quickly decays to zero at large separations, especially at $m_{\rm lat}/g=1,5$. While for lighter fermion mass as shown in Fig.~\ref{fig:corrlight}, the correlators decay much more slower and sizable tails remain around the boundary, reliable light-mass PDFs require larger systems.

Thus we use $B_1$ to estimate the property of the PDF using the finite-range correlator value. The charge conjugation symmetry ensure $f_S(-\xi)=-f_S(\xi)$ with $\xi \in [-1, 1]$, and we use the sine-weighted integral to evaluate the PDF asymmetry:
\begin{align}
 B_1&=\int_0^1\dd\xi\,f_S(\xi)\sin(2\pi\xi)\notag\\
 &=\frac{1}{2}\int_{-\infty}^{\infty}\dd\xi\,f_S(\xi)\sin(2\pi\xi)\notag\\
 &=\frac{N}{16\pi}\int_{-\infty}^{\infty}\dd\nu\,C_S(\nu)
 \int_{-\infty}^{\infty}\dd\xi\,e^{\ii\xi\nu}\sin(2\pi\xi)\notag\\
 &=\frac{N}{16\ii}\bigl[C_S(-2\pi)-C_S(2\pi)\bigr].
 \label{eq:B1C}
\end{align}
Here $f_S(\xi)=N/(8\pi)\int\dd\nu\,e^{\ii\xi\nu}C_S(\nu)$ is used in the third line. And we used
$\int\dd\xi\,e^{\ii\xi\nu}\sin(2\pi\xi)
=\pi[\delta(\nu+2\pi)-\delta(\nu-2\pi)]/\ii$.
The second line uses the oddness of $f_S$ and assumes that $f_S(\xi)$ vanishes outside $[-1,1]$.
This relation gives the correlator estimate of $B_1$ in Eq.~\eqref{eq:B1}, since
$C_S(\pm2\pi)=C_S(l_\pm^*,l_s)$, with $l_\pm^*=l_s\pm\pi/(aM_S)$.

As a cross-check at heavier fermion masses, we also evaluate the
sine moment of the average of the fermion and antifermion PDFs:
\begin{equation}
 B_1 =\int_0^1\dd\xi\,
 \frac{\operatorname{Re}[f_S(\xi)-f_S(-\xi)]}{2}\sin(2\pi\xi).
 \label{eq:B1FT}
\end{equation}

\end{document}